\documentclass{article}

\usepackage[preprint]{neurips_2024}

\usepackage{amsmath}
\usepackage{graphicx}
\usepackage[utf8]{inputenc} % allow utf-8 input
\usepackage[T1]{fontenc}    % use 8-bit T1 fonts
\usepackage{hyperref}       % hyperlinks
\usepackage{url}            % simple URL typesetting
\usepackage{booktabs}       % professional-quality tables
\usepackage{amsfonts}       % blackboard math symbols
\usepackage{nicefrac}       % compact symbols for 1/2, etc.
\usepackage{microtype}      % microtypography
\usepackage{algorithm}
\usepackage{algorithmic}
\usepackage{colortbl}
\usepackage{xcolor}
\usepackage{natbib}

\definecolor{peach}{HTML}{F26035}

\title{Super-Resolution without High-Resolution Labels for Black Hole Simulations}

\author{%
   Thomas Helfer \\
   Institute for Advanced Computational Science, Stony Brook University\\
   Stony Brook, NY 11794 USA  \\
  \texttt{thomas.helfer@stonybrook.edu} \\
   \AND
   Thomas D.P. Edwards \\
    William H. Miller III Department of Physics and Astronomy, Johns Hopkins University \\
    {(currently working at Meta)}\\
   Baltimore, Maryland 21218, USA \\
   \And
   Jessica Dafflon \\
   Valence Labs,\\
   6666 Rue Saint-Urbain \\
   Montreal, QC H2S 3H1, Canada \\
   \And
   Kaze W.K. Wong \\
    Department of Applied Mathematics and Statistics, Johns Hopkins University\\ Baltimore, MD 21218, USA\\
    Data Science and AI Institute, Johns Hopkins University 
   \\ Baltimore, MD 21218, USA\\
    Center for Computational Astrophysics, Flatiron Institute, \\
    New York, NY 10010, USA  \\
   \AND
   Matthew Lyle Olson \\
    Intel Labs \\
   Santa Clara CA, USA \\
}

\begin{document}

\maketitle

\begin{abstract}
Generating high-resolution simulations is key for advancing our understanding of one of the universe's most violent events: Black Hole mergers. However, generating Black Hole simulations is limited by prohibitive computational costs and scalability issues, reducing the simulation's fidelity and resolution achievable within reasonable time frames and resources. In this work, we introduce a novel method that circumvents these limitations by applying a super-resolution technique {\it without directly needing high-resolution labels}, leveraging the Hamiltonian and momentum constraints—fundamental equations in general relativity that govern the dynamics of spacetime. We demonstrate that our method achieves a reduction in constraint violation by one to two orders of magnitude and generalizes effectively to out-of-distribution simulations.
\end{abstract}

\section{Introduction}

\begin{figure*}[t]
\centering
\includegraphics[width=1.03\textwidth]{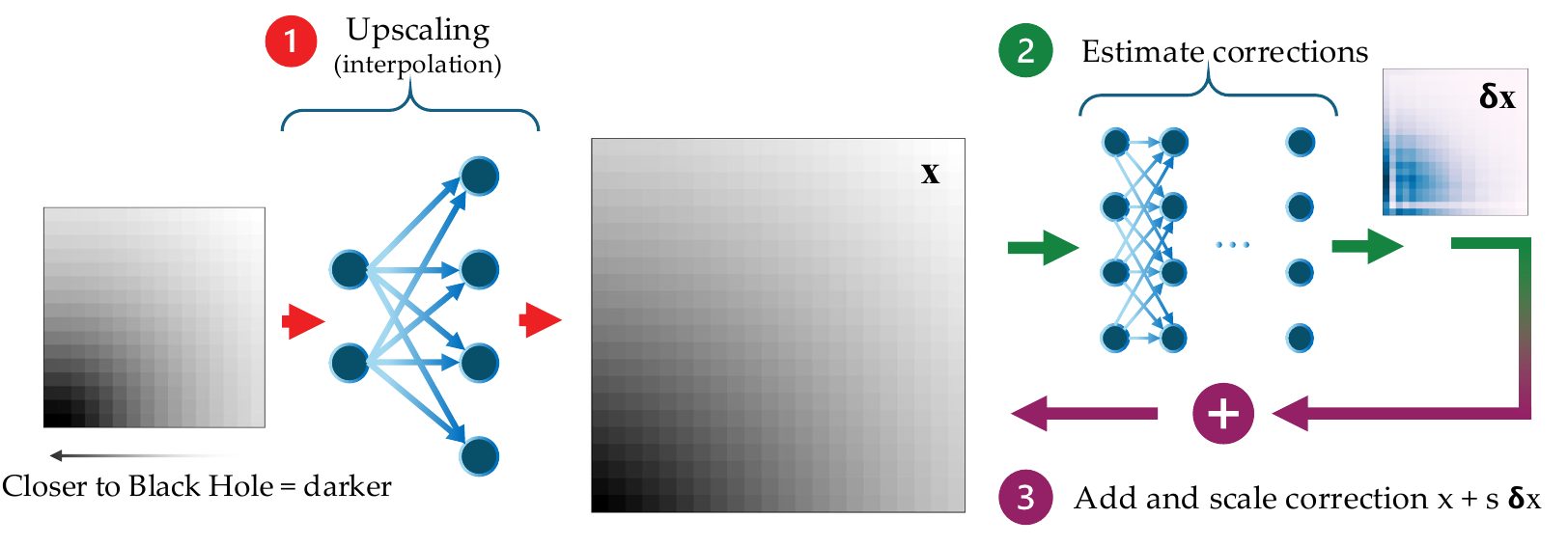}
\caption{Schematic representation of our framework: (1) We first apply a commonly used interpolation method to up-sample our simulation, then (2) a network takes the up-sampled simulation and produces a correction $\delta x$. This correction is then (3) multiplied by the scaling factor \emph{s} and added to the up-sampled simulation. The corrected simulation results in reduced constraint violations, leading to an improved simulation.}\label{fig:schematic}
\end{figure*}

The advent of gravitational wave astronomy has started a new era in astrophysics, enabling new insights into some of the universe's most violent events, such as Black Hole mergers and neutron star collisions \citep{LIGOScientific:2018mvr,LIGOScientific:2020ibl,LIGOScientific:2021usb,KAGRA:2021vkt}. Numerical relativity (NR) simulations play a crucial role in predicting the waveforms of such phenomena and are essential for the successful analysis of data observed by gravitational wave detectors like LIGO and Virgo \citep{LIGOScientific:2016sjg}.
As the sensitivity of upcoming detectors (e.g., LISA \citet{amaroseoane2017laserinterferometerspaceantenna}) will increase by orders of magnitude, the demand for more accurate and diverse waveforms generated by NR simulations grows exponentially \citep{LISAConsortiumWaveformWorkingGroup:2023arg}. However, existing numerical methods face challenges in meeting these demands, which could limit the scientific return on the significant financial investments made in these detectors.

Next-generation detectors will require more advanced solutions capable of handling longer, higher-resolution simulations and larger mass ratios for Black Hole binaries. In this work, we present a super-resolution-inspired method that employs a convolutional neural network and uses constraints from general relativity to make the network physics-aware. The method is designed to be applied to current state-of-art numerical codes and aims to reduce simulation error and enhance the accuracy of gravitational waveform predictions.

Our framework aims to enhance adaptive mesh refinement (AMR), a widely used technique for dynamically adjusting resolution in simulations. AMR is particularly effective in situations where only specific regions, such as near black hole horizons, require higher resolution, while the rest of the simulation can operate at lower resolution. However, transitioning from low- to high-resolution grids requires the use of higher-order interpolation methods \citep{Adams2014ChomboSP}, which can introduce errors. To reduce these errors, we developed a super-resolution method that can be used to improve the overall accuracy of AMR simulations. 

Most super-resolution techniques, however,  require high-resolution labels for the training. Getting these high-resolution labels for us requires expensive simulations, and to avoid this computational cost, we propose a framework that uses a unique loss function derived from general relativity's physical constraints. These constraints -- referred to as Hamiltonian and Momentum constraints -- are used for  monitoring the stability of simulations. If they are not fulfilled below a threshold or show fast-growing trends, it is a strong indication of a problem in the simulation. 

Another improvement brought by our framework is the ability to harness the power of parallel processing via GPUs, offering improved computational efficiency and scalability. As modern NR codebases evolve, there is a growing shift towards incorporating hardware accelerators like GPUs, in contrast to the traditionally CPU-based compute environments. This transition of NR codebases is critical, as an increasing portion of available computational resources have GPUs. 
Furthermore, this work demonstrates a novel application on how to leverage deep learning for numerical relativity. We believe that deep learning offers numerous opportunities to enhance NR, and when applied correctly, it can help close the gap in numerical performance for the next-generation gravitational wave detectors. 

Our paper is organized as follows. In Section 2, we provide a brief overview of numerical relativity. In Section 3, we introduce our super-resolution framework, describing the neural network architecture, loss functions, and data generation process. In Section 4, we present our results, demonstrating the performance and generalizability of our framework, with an in-depth comparison to the $L_1$ loss using ground truth data. We conclude in Section 5, highlighting key results and discussing the limitations of this work.

\begin{figure}[t]
\centering
\includegraphics[width=0.5\linewidth]{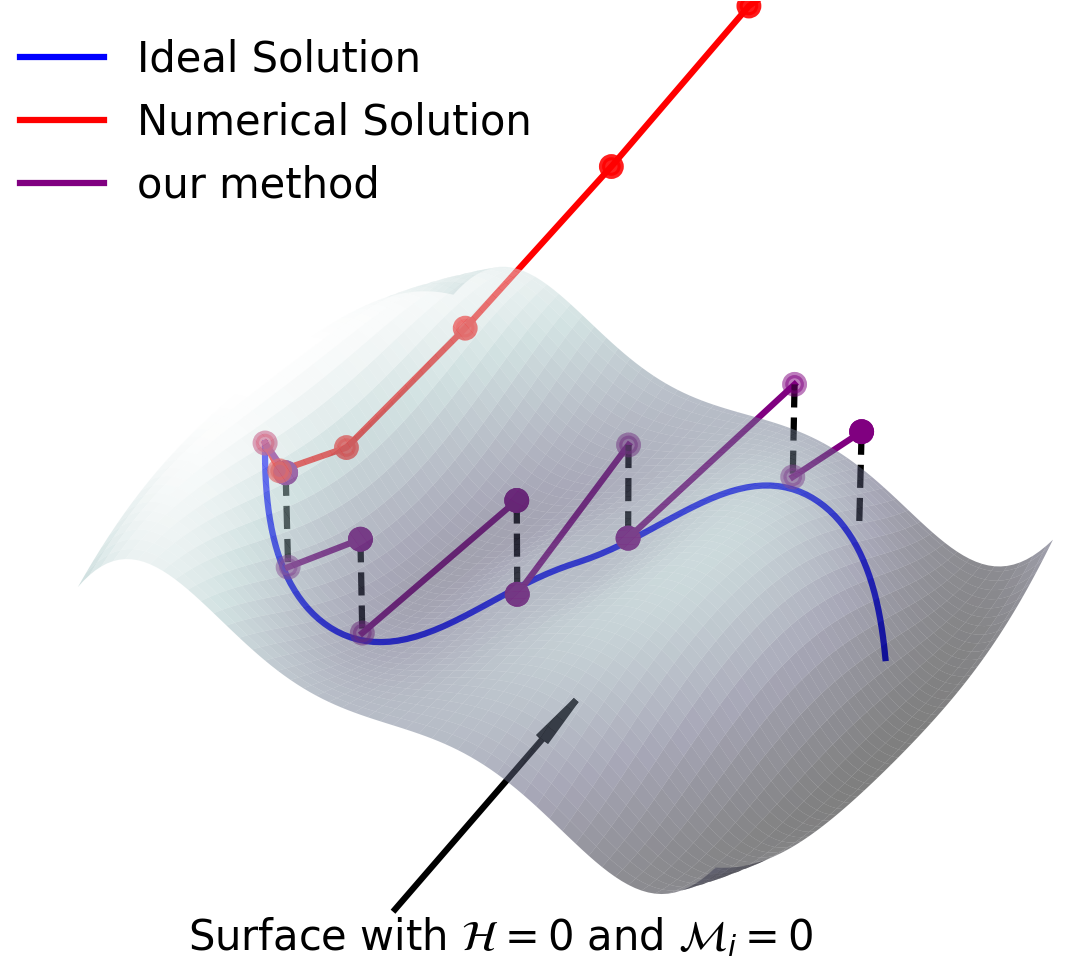}
\caption{Visualization of the solution surface with $\mathcal{H} = 0$ and $\mathcal{M}_i = 0$. The blue line represents the ideal solution when following Einstein's equations, while the red line illustrate a numerical solution. The dots on both the red and purple line represent the discretization of time. Our method (purple) projects the solution back to the surface where constraint are fulfilled to produce a numerical result closer to the ideal solution.}
\label{fig:manifold}
\end{figure}

\section{Background: Numerical Relativity}

Numerical relativity (NR) provides the computational framework for simulating the complex dynamics of spacetime, such as those observed in Black Hole mergers and gravitational waves. This section outlines the core concepts of numerical relativity, offering a small overview of its theoretical underpinnings. For those interested in a more in-depth exploration, please refer to \cite{baumgarte2021numerical,baumgarte2010numerical,alcubierre2008introduction}.

We work in natural units where the speed of light $c = 1$ and the gravitational constant $G_N = 1$ to simplify equations. Throughout this work, we employ Einstein summation notation, which simplifies expressions involving summations over indexed variables in tensor algebra. This notation implies that any index appearing twice in a single term is summed over. For example, the summation
$
C_\mu = \sum_{\nu=1}^n A_{\mu\nu} B^\nu,
$
where $n$ is the number of elements of the tensor, can be written concisely using Einstein notation as $C_\mu = A_{\mu\nu} B^\nu$. Additionally, a raised index indicates summation with the inverse of the spacetime metric tensor $g_{\mu\nu}$, which we express as $g^{\mu\nu}$. For instance, 
$
A^\mu = g^{\mu\nu} A_\nu.
$
When using Latin indices ($i, j, k, \dots$) in place of Greek indices ($\alpha, \beta, \gamma, \dots$), we employ the projected spatial metric tensor $\gamma^{ij}$ to raise and lower indices in the three-dimensional spatial subspace.

\begin{table}[t]
\centering
\caption{{\bf Simulation data } parameters used for training and test. Each simulation has multiple resolution levels, each with a different grid spacing d$x$. Furthermore, we give the number of boxes and how many numerical time steps the simulation performed on each level. Due to our choice of loss, we need to train a different model for each level. }
\begin{tabular}{c|c|c|c|c }
\hline
resolution level & size & grid spacing d$x$ & number of boxes & time-steps  \\
\hline
%(64.0 * 2**res_level) / 512.0 
 5 & 263M & 0.25 &336 & 7 \\
 6 & 938M & 0.125 &1200 & 15 \\
 7 & 5.2G & 0.0625 &6696 & 31  \\ 
 8 & 19G & 0.03125 &24704 & 64 \\
 9 & 15G & 0.015625 &19208 & 99\\\hline
\end{tabular}
\label{table:overview_res_level_size}
\end{table}

\subsection{Foundations of General Relativity}

The theoretical backbone of NR is Einstein's general theory of relativity \citep{einstein1916foundation}, which is described by the equation 
\begin{equation}\label{eqn:Einsteinsfieldsequation}
G_{\mu\nu} = 8\pi T_{\mu\nu}.
\end{equation}
This equation describes how matter and energy (encoded in the stress-energy tensor $T_{\mu\nu}$) influence the curvature of spacetime, represented by the Einstein tensor $G_{\mu\nu}$. 

The direct application of Einstein's equations in their original form is not feasible in NR simulations due to a lack of distinction between time and space. This challenge is addressed by the Arnowitt-Deser-Misner (ADM) (3+1) decomposition \citep{PhysRev.116.1322}, a mathematical formalism that reformulates Einstein's equations into a set suitable for numerical analysis. To be also numerically stable, we use the standard CCZ4 formulation \citep{PhysRevD.88.064049,PhysRevD.85.064040}.

Einstein's equation in the ADM decomposition gives rise to the constraint equations that we propose for our framework as following
%After some derivation we obtain the following equations:
\begin{equation}
    \mathcal{H} := R + \frac{2}{3}K^2 
    - \tilde{A}_{kl}\tilde{A}^{kl}-16\pi\rho,
    \label{eq:Ham-constraint}\\
\end{equation}
\begin{equation}
    %     \begin{split}
        \mathcal{M}_i := \tilde{\gamma}^{kl}\left(\partial_k \tilde{A}_{li} 
        -2\tilde{\Gamma}^m_{l(i}\tilde{A}_{k)m}
        -3\tilde{A}_{ik}\frac{\partial_l\chi}{2\chi}\right)
%         \\&\qquad\qquad\qquad\qquad\qquad
        -\frac{2}{3}\partial_iK - 8\pi S_i,
%     \end{split}
    \label{eq:mom-constraint}
\end{equation}

where ${R}$ is the Ricci scalar, $\chi$, $K$, $\tilde{A}_{ik}$ and $\tilde{\gamma}^{kl}$ are evolution variables, $\rho$ and $S_i$ are energy and momentum density -- both describe matter moving on the space manifold (e.g., Neutron Stars, Humans, Photons, cats). These equations $\mathcal{H}$ and  $\mathcal{M}_i$ need to be equal to zero to be consistent with general relativity. However, this is never truly possible in numerical methods as the discretization introduces small errors. Although there are many methods that try to minimize this error by modifying the evolution equations (as was done in the CCZ4 formulation), with our framework, we introduce a physics-aware network to minimize these errors.

\section{Methodology}

Our approach combines deep learning and NR techniques to improve Black Hole simulations. We designed a super-resolution framework that applies a neural network correction to up-sampled simulations using higher order interpolation used in the GRTL code, optimizing it with a physics-aware loss based on the Hamiltonian and Momentum constraints of general relativity. By performing small correction to enforce constraints, we aim to achieve a more accurate simulation as depicted in Fig.~\ref{fig:manifold}.
The code used for training and evaluation is available at \url{https://github.com/ThomasHelfer/TorchGRTL}. The higher order interpolation is separately available at \url{https://github.com/ThomasHelfer/PyInterpX} and via \texttt{pip install pyinterpx}.

\subsection{Loss functions}

In contrast to supervised ML methods -- where we would use the distance between predicted and ground truth as a loss -- here we use the sum of squares of the violation of Eq.~\ref{eq:Ham-constraint} and Eq.~\ref{eq:mom-constraint}. 
\begin{equation}
    \mathcal{L}_{\rm GR} =  \sum_j \left(|\mathcal{H}(\mathbf{x}_j)|^2 + \sum_{i = 0}^D |\mathcal{M}_i(\mathbf{x}_j)|^2 \right),
\end{equation}
where D is the number of spatial dimensions (three for our experiments) and $\mathbf{x}_j$ represent positions on the simulation grid. We defined the normalized loss to evaluate the performance of our model and the baseline obtained by the higher order interpolator currently used in some NR codes as
\begin{equation}
    {\rm Normalized\, \mathcal{L}_{\rm GR}} = \frac{\mathcal{L}_{\rm GR}({\rm Our\, method })}{\mathcal{L}_{\rm GR} ({\rm Baseline})}\label{eq:normalise_performance}.
\end{equation}

\begin{figure*}[t]
\centering
\includegraphics[width=1.02\textwidth]{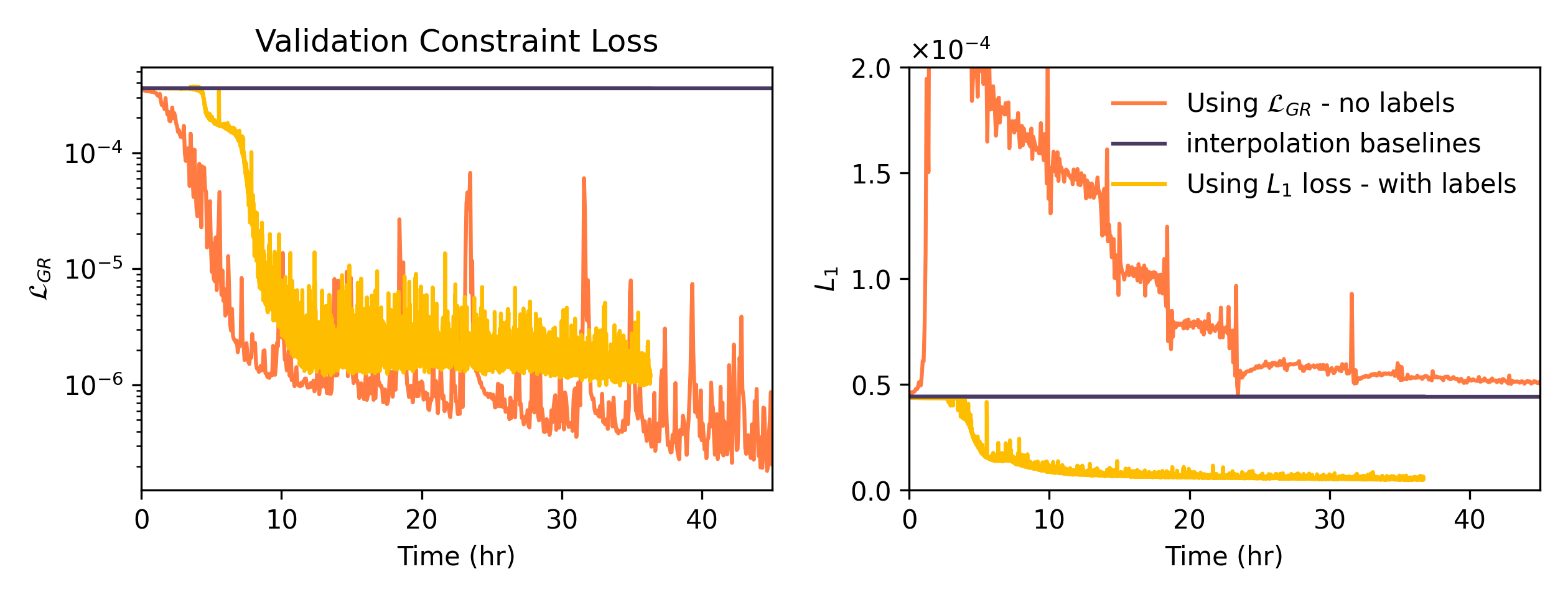}
\caption{Comparing the performance of models trained with ${L_1}$ loss (which requires high-resolution labels) and $\mathcal{L}_{\rm GR}$ loss (which does not require high-resolution labels). As shown in the left figure, both ${L_1}$ and $\mathcal{L}_{\rm GR}$ losses converge to similarly small values of $\mathcal{L}_{\rm GR}$. However, in the right figure, we observe that both $\mathcal{L}_{\rm GR}$ and $L_1$ converge to different values of $L_1$. This suggests that the two losses lead to different solutions, likely due to the under-specified nature of the constraint-based $\mathcal{L}_{\rm GR}$ loss. The dark straight line is the interpolation used in GRTL code.
}\label{fig:twosolutionsLoss}
\end{figure*}
\subsection{Dataset generation} Our framework employs GRTeclyn (formerly GRChombo \citep{Radia:2021smk,Andrade:2021rbd}) an established open-source codebase for the NR simulations.
We simulate two equal-mass Black Holes in a quasi-circular orbit\footnote{Data is available at \url{https://huggingface.co/datasets/thelfer/BinaryBlackHole}.}, a common reference benchmark in numerical relativity (similar to how the MNIST dataset is used as a benchmark in computer vision).

To be able to fit our simulations in GPU memory we subdivided them into blocks of $(16 \times 16 \times 16)$ points and 25 channels representing different evolution variables (i.e., $\chi$, $K$, $\tilde{A}_{ik}$, $\tilde{\gamma}_{kl}$, $\alpha$ and $\beta^i$). We trained on 80\% of the data and used 20\% to test the in-distribution performance. To test out-of-distribution performance, we also created several independent simulations with increasing Black Hole masses\footnote{Data is available at \url{https://huggingface.co/datasets/thelfer/BinaryBlackHoleValidation}} (See Table \ref{table:out-of-dist} for details).

In our adaptive mesh simulations, we work with multiple resolution levels, each generating data at different rates. Due to the nature of our loss function, we need to train separate models for each resolution level. Notably, the amount of data varies between these levels,
ranging from 263 MB to 19 GB (See Table \ref{table:overview_res_level_size}). The highest resolution levels typically generate the most data because our simulation uses subcycling, where higher levels undergo more time steps relative to the lower levels.

\subsection{Framework architecture}\label{sec:Framework}

An overview of the framework architecture can be found in Fig.~\ref{fig:schematic}. First we up-sample our simulation from low resolutions using a higher order interpolator commonly used throughout Black Holes simulations \citep{Schnetter:2003rb}. The upsampled simulations will not only be the input for the neural network that calculates the correction ${\delta} x$, but also serve as the baseline. As we do not know the scale of the correction a priori, we introduce a rescale factor $s$ to reduce floating-point problems. So, our correction is
\begin{equation}
    x +  s \delta x ~,\label{eqn:corr}
\end{equation}
where $x$ is the vector of all variables $x = (\chi, K, \tilde{A}_{ik}, \dots)$ and for our data, we found that  $s = 10^{-4}$ or $10^{-5}$ works well.

\begin{figure}
\centering
\includegraphics[width=0.63\linewidth]{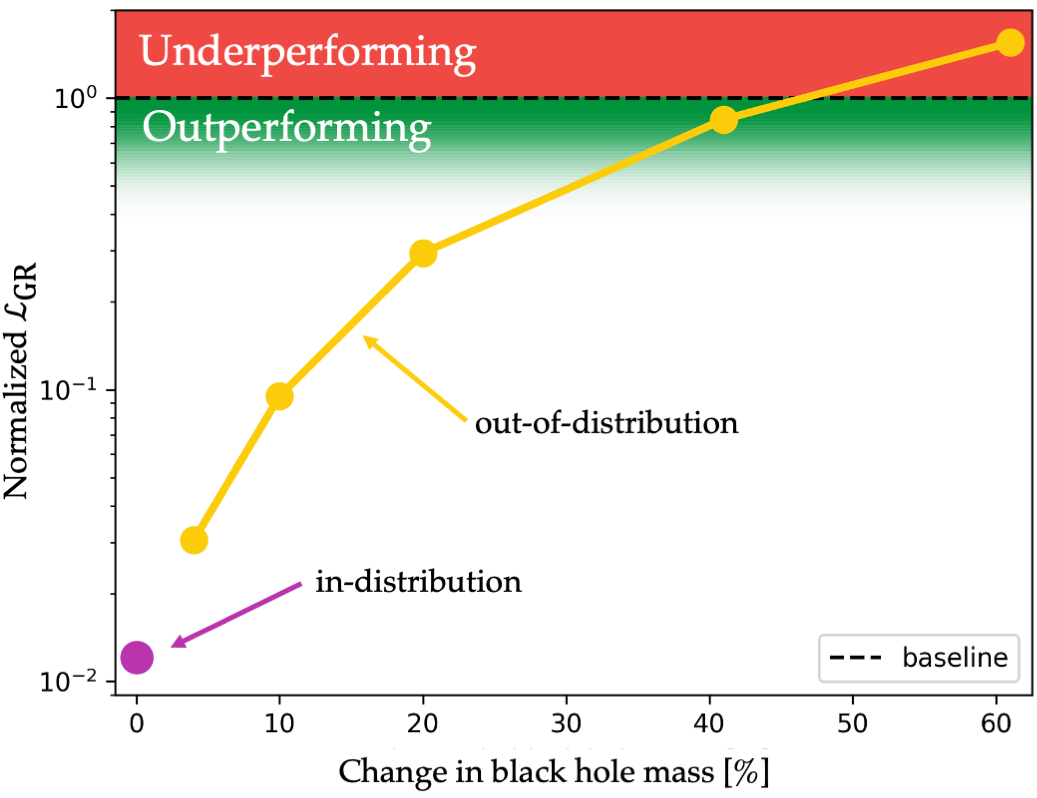}
\label{fig:Money_fig}
\caption{\textbf{Our framework (purple and yellow) outperforms the baseline (dotted black) by two orders of magnitude}. In NR simulations, the mass of a Black Hole is a parameter for defining the simulation. We evaluate the loss of the validation set in the in-distribution scenario (purple). However, we aimed to stress-test our framework by varying the Black Hole's mass, enabling us to \textit{evaluate its ability to generalize to out-of-distribution scenarios} (yellow). Remarkably, even with a 41\% variation in the Black Hole's mass, our framework still outperforms the baseline. A more complete overview can be found in Tabel \ref{tab:Money_table}.}
\end{figure}

\paragraph{Numerical Stability}
As the Hamiltonian and Momentum constraints (Eq.~\ref{eq:Ham-constraint} and \ref{eq:mom-constraint}) are mathematically underspecified (we have 25 variables and only four equations), there are many possible solutions that the system can take. We can address this by introducing masking, changing our correction (Eq.~\ref{eqn:corr}) at training to 
\begin{equation}\label{eqn:masking}
    x +  m \cdot s\delta x ~,
\end{equation}
where $m$ is a mask. We found that a mask where every fourth element in each direction is masked, with a random shift applied, performed best in guiding the solution compared to a fully random mask.

To clarify the effectiveness of this approach, consider a scenario where $\delta x$ is large. When comparing two neighboring points, if one is masked while the other is not, there is a significant difference between them. This difference leads to large gradients, which increases the $\mathcal{L}_{\rm GR}$ loss. As a result, solutions with large $\delta x$ are penalized. By encouraging smaller values of $\delta x$, we keep the corrected simulation close to its original state $x$.

To gain further insights into potential solutions within our framework, we conducted experiments by down-sampling boxes with \(16 \times 16 \times 16\) resolution down to \(8 \times 8 \times 8\), allowing us to establish a clear ground truth for evaluating our up-sampling methods. To maintain comparability, we aligned the interpolation grid such that every second element is directly copied from the low-resolution data (see Fig. \ref{fig:gridorientation}), while the other elements are determined through interpolation.

\paragraph{Neural Network Details}
Since translation symmetry is inherently encoded in Einstein's gravity when using a Cartesian grid, convolutional neural networks are a natural choice for our framework. We constructed a convolutional neural network composed of 4 hidden layers with 64 channels and ReLU non-linearity at the end of every hidden layer. 
To keep both input and output sizes the same, we employ padding. Lastly, as the gradients involved in this process can be very small, we use double precision to avoid any underflow issues.

\section{Results}
Our framework improves simulation quality by one to two orders of magnitude, as measured by our loss $\mathcal{L}_{\rm GR}$, with performance varying based on the specific data and model. It also generalizes well to out-of-distribution simulations, as shown in Fig.~\ref{fig:Money_fig} and Table \ref{tab:Money_table} for a comprehensive overview. The important contribution of our paper is the introduction of the physics-aware loss $\mathcal{L}_{\rm GR}$, because of it our approach does not need any high resolution labels, making the data generation significantly cheaper. While the $\mathcal{L}_{\rm GR}$ loss incurs a higher computational cost per iteration compared to the $L_1$, it converges significantly faster. In our experiments, the slower per-iteration time is roughly balanced by fewer required iterations, resulting in similar overall convergence times compared to $L_1$ (See Fig.~\ref{fig:twosolutionsLoss}).

There are two ways to use the presented framework (see Table~\ref{tab:Money_table}). The first, as discussed so far, is to correct errors introduced by the interpolation routine in adaptive mesh refinement. The second approach involves maintaining the original resolution of the simulation and using the framework to correct numerical errors that accumulate during the simulation. This second use case extends the framework’s applicability beyond adaptive mesh refinement. By applying it directly to fixed-resolution or single-resolution simulations, our method can enhance simulations on any mesh. This capability is particularly valuable for improving the accuracy of complex simulations prone to constraint buildup, such as those in modified gravity scenarios or simulations of highly spinning black holes. 

\begin{figure}[t]
\centering
\includegraphics[width=0.45\textwidth]{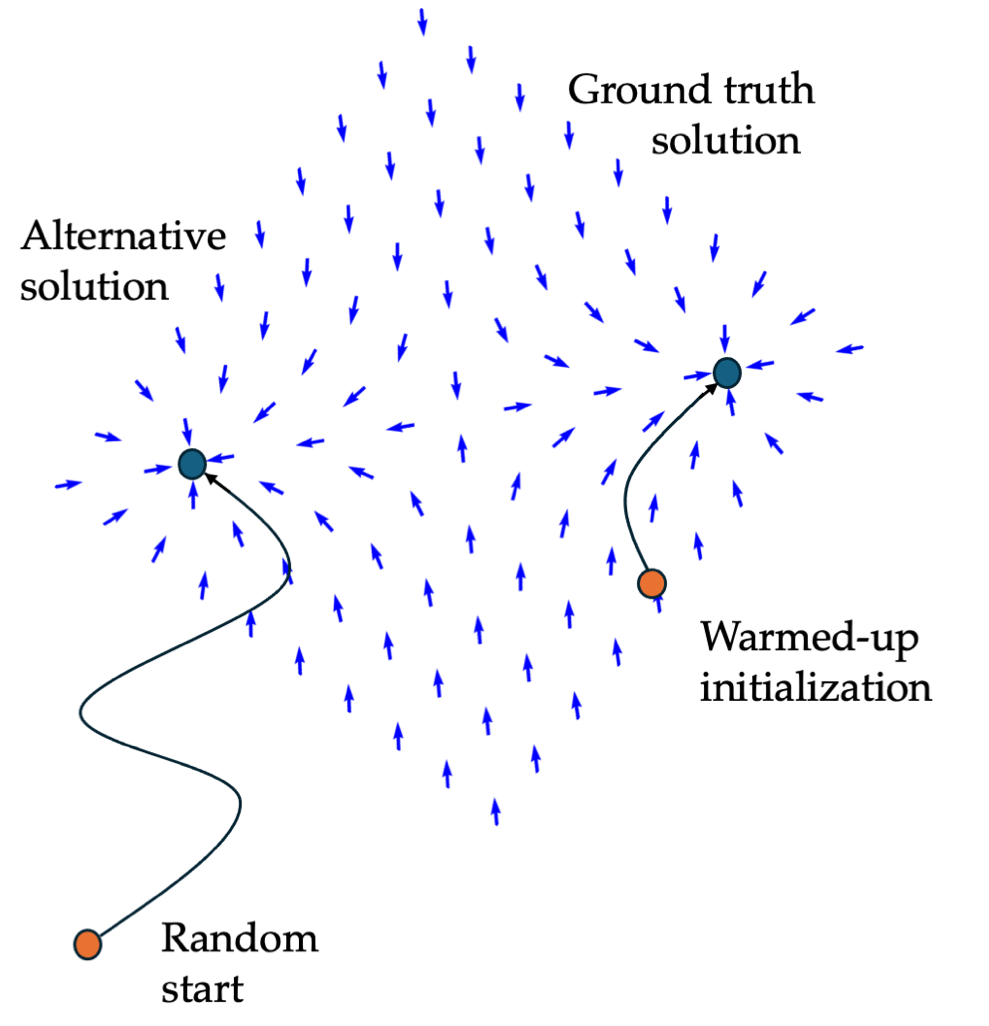}
\caption{{\bf Multiple solutions:} Depending on starting point, we can control what state our simulations falls into when using the $\mathcal{L}_{\rm GR}$ loss. When we start with a pre-trained network (marked as Warmed-up initialization), we can converge towards the ground truth.}\label{fig:flowstate}
\end{figure}

\subsection{Multiple Solutions and how to control them}\label{sec:mul_res}

In this section, we compare the results of training with the $L_1$ loss versus the \(\mathcal{L}_{\rm GR}\) loss. In Fig.~\ref{fig:twosolutionsLoss} (left), we observe that training with the $L_1$ loss also minimizes \(\mathcal{L}_{\rm GR}\), which is expected since the ground truth approximately satisfies the $\mathcal{L}_{\rm GR}$ constraints. However, Fig.~\ref{fig:twosolutionsLoss} (right) shows that training with the $\mathcal{L}_{\rm GR}$ loss versus the $L_1$ loss leads to different solutions. When trained with the \(\mathcal{L}_{\rm GR}\) loss, the model converges to a solution close to the interpolation baseline. In contrast, training with the $L_1$ loss produces a solution closer to 0, indicating alignment with the ground truth. These differences in solutions are due to the under-specified nature of the $\mathcal{L}_{\rm GR}$ loss, which allows for multiple solutions that satisfy the constraints but are distinct from the ground truth.

\begin{figure}[t]
\centering
\includegraphics[width=1.03\textwidth]{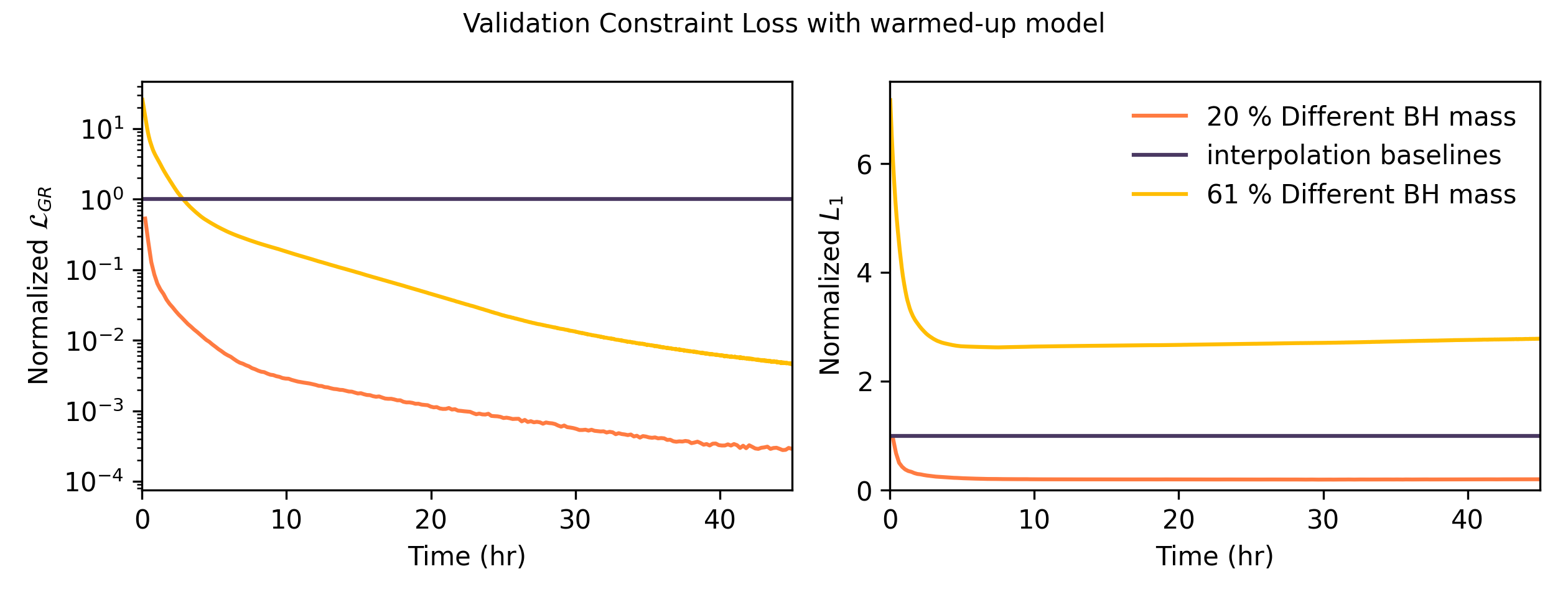}
\caption{The effect of using a warmed-up model. In this figure, we use a network pre-trained with a supervised $L_1$ loss and fine-tune it using the $\mathcal{L}_{\rm GR}$ loss on a different dataset. This is done on two datasets: one where the binary black hole mass differs by 20\% and another where it differs by 61\%. Both cases show convergence in $\mathcal{L}_{\rm GR}$ (in the left figure). The first (with 20\% mass difference) approaches zero in normalized $L_{1}$ (right figure), indicating that the $\mathcal{L}_{\rm GR}$ loss can converge toward the ground truth if initialized properly. However, the 61 \% network is too different and thus does not converge close to the ground truth.}
\label{fig:warmed-up-loss}
\end{figure}

% Experiment name silver-tree-304
It is also possible to achieve convergence to the ground truth solution without directly using the $L_1$ loss (as depicted in Fig.~\ref{fig:flowstate}). This is based on the observation that the ground truth solution found by $L_1$ is, at least approximately, also a solution for $\mathcal{L}_{\rm GR}$. By initializing closer to the desired solution, we can improve convergence to that solution under $\mathcal{L}_{\rm GR}$. To demonstrate this, we took a network already trained with $L_1$ loss and resumed training using $\mathcal{L}_{\rm GR}$ on a different simulation, where the mass differed by roughly 20\% relative to the initial simulation used for $L_1$ training. We refer to this as the 'warmed-up' solution. We observed that this differently initialized network converged to a state with a small $\mathcal{L}_1$ value (See Fig.~\ref{fig:warmed-up-loss}). However, if the difference in mass becomes too large, we are no longer able to achieve this convergence.

In this paper, we do not claim that one solution is inherently preferable for numerical simulations. However, if a solution close to the ground truth is desired, our experiments suggest that a combination of training strategies can help achieve this without fully relying on high-resolution labels. Unlike typical foundation models, which often involve pre-training on large unlabeled datasets followed by fine-tuning on labeled data, we propose a reverse approach: pre-training on a small labeled dataset with $L_1$ loss, followed by post-training on a larger, cheaper unlabeled dataset using $\mathcal{L}_{\rm GR}$ loss.

\begin{table*}
    \centering
    \caption{{\bf The normalized loss} of models in our GitHub repository (\url{https://github.com/ThomasHelfer/TorchGRTL/tree/main/models}). The column ``config-name'' corresponds to the folder name in the repository, ``factor'' specifies the upscaling factor applied to the simulation, and ``level'' indicates the resolution level at which the model was trained. A factor of 1 corresponds to the use case where the framework is applied at the original simulation resolution to correct accumulated numerical errors, while factors of 2 and 4 represent the use case where the framework corrects errors introduced by the interpolation routine. For better visualization, we plotted data from the model ``x1-lvl9'' in Fig.~\ref{fig:Money_fig}. Lastly, $\Delta M = 0$ represents the test set from the training simulation, while other values of $\Delta M$ indicate our out-of-distribution datasets.
}
    \begin{tabular}{lcc|cccccc}
    \toprule
    \toprule
      & & & \multicolumn{6}{c}{change in Black Hole mass $\Delta M \rightarrow $} \\ \cmidrule{4-9}
     config-name & factor &  level & $ 0 \% $ & $ 4 \% $  & $ 10 \% $ & $ 20 \% $ & $ 41 \% $ & $ 61 \% $ \\ 
     \midrule
x1-lvl5 & 1 & 5 & \cellcolor{green!58}{0.172} & \cellcolor{green!100}{0.090} & \cellcolor{green!100}{0.090} & \cellcolor{green!99}{0.100} & \cellcolor{green!48}{0.205} & \cellcolor{red!20}{2.078} \\
x1-lvl6 & 1 & 6 & \cellcolor{green!100}{0.067} & \cellcolor{green!54}{0.183} & \cellcolor{green!42}{0.234} & \cellcolor{green!23}{0.427} & \cellcolor{red!13}{1.361} & \cellcolor{red!45}{4.570} \\
x1-lvl7 & 1 & 7 & \cellcolor{green!69}{0.144} & \cellcolor{green!60}{0.166} & \cellcolor{green!43}{0.228} & \cellcolor{green!20}{0.488} & \cellcolor{red!19}{1.971} & \cellcolor{red!53}{5.378} \\
%x1-lvl8 & 1 & 8 & \cellcolor{green!10}{0.996} & \cellcolor{green!10}{0.996} & \cellcolor{green!10}{0.995} & \cellcolor{green!10}{0.994} & \cellcolor{green!10}{0.993} & \cellcolor{green!10}{0.992} \\
x1-lvl9 & 1 & 9 & \cellcolor{green!100}{0.012} & \cellcolor{green!100}{0.030} & \cellcolor{green!100}{0.095} & \cellcolor{green!33}{0.295} & \cellcolor{green!11}{0.844} & \cellcolor{red!15}{1.549} \\
x2-lvl5 & 2 & 5 & \cellcolor{green!66}{0.150} & \cellcolor{green!43}{0.232} & \cellcolor{green!23}{0.420} & \cellcolor{red!13}{1.345} & \cellcolor{red!58}{5.838} & \cellcolor{red!100}{15.154} \\
x2-lvl6 & 2 & 6 & \cellcolor{green!100}{0.060} & \cellcolor{green!67}{0.148} & \cellcolor{green!32}{0.312} & \cellcolor{red!11}{1.129} & \cellcolor{red!61}{6.140} & \cellcolor{red!100}{19.088} \\
x2-lvl7 & 2 & 7 & \cellcolor{green!100}{0.077} & \cellcolor{green!82}{0.121} & \cellcolor{green!30}{0.324} & \cellcolor{red!12}{1.242} & \cellcolor{red!72}{7.280} & \cellcolor{red!100}{23.514} \\
x2-lvl8 & 2 & 8 & \cellcolor{green!63}{0.158} & \cellcolor{green!53}{0.187} & \cellcolor{green!30}{0.326} & \cellcolor{red!10}{1.019} & \cellcolor{red!44}{4.450} & \cellcolor{red!100}{13.110} \\
x2-lvl9 & 2 & 9 & \cellcolor{green!43}{0.231} & \cellcolor{green!30}{0.326} & \cellcolor{green!13}{0.749} & \cellcolor{red!14}{1.483} & \cellcolor{red!26}{2.608} & \cellcolor{red!57}{5.771} \\
x4-lvl5 & 4 & 5 & \cellcolor{green!17}{0.585} & \cellcolor{green!18}{0.546} & \cellcolor{green!16}{0.612} & \cellcolor{green!13}{0.764} & \cellcolor{red!13}{1.390} & \cellcolor{red!23}{2.383} \\
    \bottomrule
    \end{tabular}

    \label{tab:Money_table}
\end{table*}

\subsection{Related work}

It is important to draw a clear distinction between Physical Informed Neural Networks (PINNs \citet{raissi2017physics}), which also use partial differential equations (PDEs) as a loss. While our framework takes an approximation of the solution as input and uses the physical constraints to improve it, PINNs take the spatial coordinates and produce the value of the PDE solution at the given coordinate. Furthermore, while PINNs need to be retrained for each new simulation, our method needs to be trained once and can then be inferred on different simulations. 

\subsection{Limitations}

Despite the improvements that our method brings, there are three main limitations that could improve the impact of our analysis. Firstly, our analysis is conducted offline,  our framework is applied only after the simulation has finished. Ideally, our framework would be applied online, replacing the baseline interpolation within an adaptive mesh codebase. However, implementing an online approach presents additional challenges, such as the engineering effort required to integrate our method with a NR codebase. These hurdles will be addressed in future research. Secondly, the trained neural network depends on the grid spacing (${\rm d} x$). Therefore, to operate across the varying spacings present in an adaptive mesh solver, with our current methods training a separate network for each resolution would be required. Thirdly, we are not fully using all available symmetries in our system. We expect that including equivariance with respect rotational symmetries \citep{gregory2023geometricimagenetextendingconvolutionalneural} would result in reduced need of training data.

\section{Conclusion}
In this work, we tested the applicability of deep learning techniques to numerical relativity simulations. We demonstrated that by introducing a physics-aware loss, we enable solutions without high-resolution labels, or, if a ground truth solution is preferred, require only a small amount of high-resolution data to guide the network. This significantly reduces reliance on computationally expensive data.

\begin{ack}
Research reported in this publication was supported by a Postdoctoral Fellowship at the Institute for Advanced Computational Science, Stony Brook University. This work was carried out at the Advanced Research Computing at Hopkins (ARCH) core facility  (rockfish.jhu.edu), which is supported by the National Science Foundation (NSF) grant number OAC1920103. The Flatiron Institute is a division of the Simons Foundation.
\end{ack}

\bibliographystyle{plainnat}
\bibliography{mybib}

\appendix

\begin{table}
\caption{Summary of masses of Black Holes in the binary simulation generated for training and out-of-distribution (OOD) performance evaluation. The table lists corresponding Black Hole masses ($M$) and percentage increases ($\Delta M$) relative to the training mass.}
\centering
\begin{tabular}{|c|c|c|c | }
\hline
Black Hole Mass $M$ & $\Delta M$ &  used for \\
\hline
0.4884 & 0 & training\\
0.5084 & 4\% & OOD\\ 
0.5384 & 10\% & OOD\\
0.5884 & 20\% & OOD\\
0.6884 & 41\% & OOD\\
0.7884 & 61\% & OOD\\\hline
\end{tabular}

\label{table:out-of-dist}
\end{table}

\begin{figure}[t]
\centering
\includegraphics[width=0.45\textwidth]{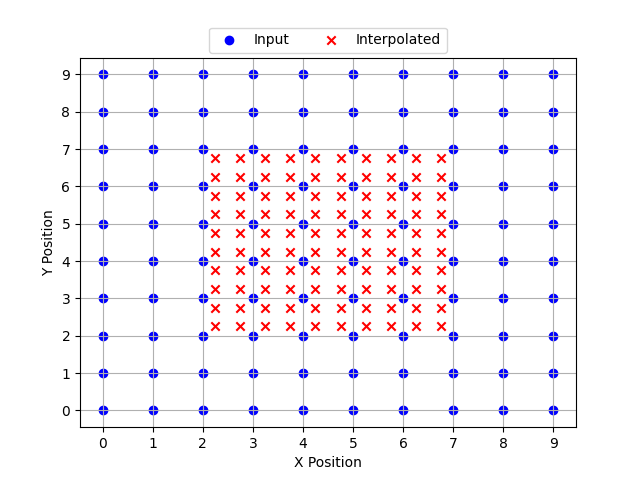}
\includegraphics[width=0.45\textwidth]{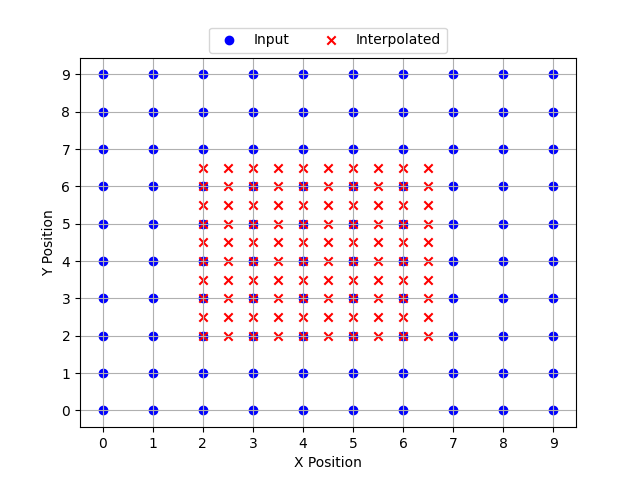}
\caption{Different grid alignments are used throughout the paper for interpolation. The blue points represent the low-resolution grid, while the red points correspond to the high-resolution interpolation grid. The left image shows the alignment used in the GRTeclyn code, which is applied to all models in Table~\ref{tab:Money_table}. The image on the right demonstrates the alignment chosen for the experiment in Section~\ref{sec:Framework}, where downsampling is simplified by selecting every second element. 
}\label{fig:gridorientation}
\end{figure}

\end{document}